\documentclass[journal=jpclcd,manuscript=letter]{achemso}
\usepackage[version=3]{mhchem}
\usepackage{xcolor}
\usepackage[utf8]{inputenc}
\usepackage[T1]{fontenc}

\author{Muhammad Nawaz Qaisrani}

\email{muhammad-nawaz.qaisrani@tu-ilmenau.de}

\affiliation{
    Ilmenau University of Technology, Theoretical Solid State Physics\\ Weimarer Straße 32, 98693 Ilmenau, Germany
}

\alsoaffiliation{
    ICTP - The Abdus Salam International Centre for Theoretical Physics\\
    Strada Costiera 11, 34151, Trieste, Italy
}

\author{Nandha Kumar}
\affiliation{
    Department of Biomaterials (Prosthodontics), Saveetha Dental College and Hospitals, Chennai, Tamil Nadu 600077, India
}

\alsoaffiliation{
    ICTP - The Abdus Salam International Centre for Theoretical Physics\\
    Strada Costiera 11, 34151, Trieste, Italy
}

\author{Christian Dreßler}
\affiliation{
    Ilmenau University of Technology, Theoretical Solid State Physics\\ Weimarer Straße 32, 98693 Ilmenau, Germany
}

\author{Ralph Gebauer}
\affiliation{
    ICTP - The Abdus Salam International Centre for Theoretical Physics\\
    Strada Costiera 11, 34151, Trieste, Italy
}

\author{Ali Hassanali}
\affiliation{
    ICTP - The Abdus Salam International Centre for Theoretical Physics\\
    Strada Costiera 11, 34151, Trieste, Italy
}

\email{ahassana@ictp.it}

\title{Acid Base Chemistry of Short Hydrogen Bonds: A Tale of Schrödinger's Cat in Glutamine Derived Crystals}

\abbreviations{PT, SHB, NQE, AIMD, PI-AIMD, WCs}


\begin{document}

\begin{tocentry}
\centering
\includegraphics[width=5cm, height=3.9cm]{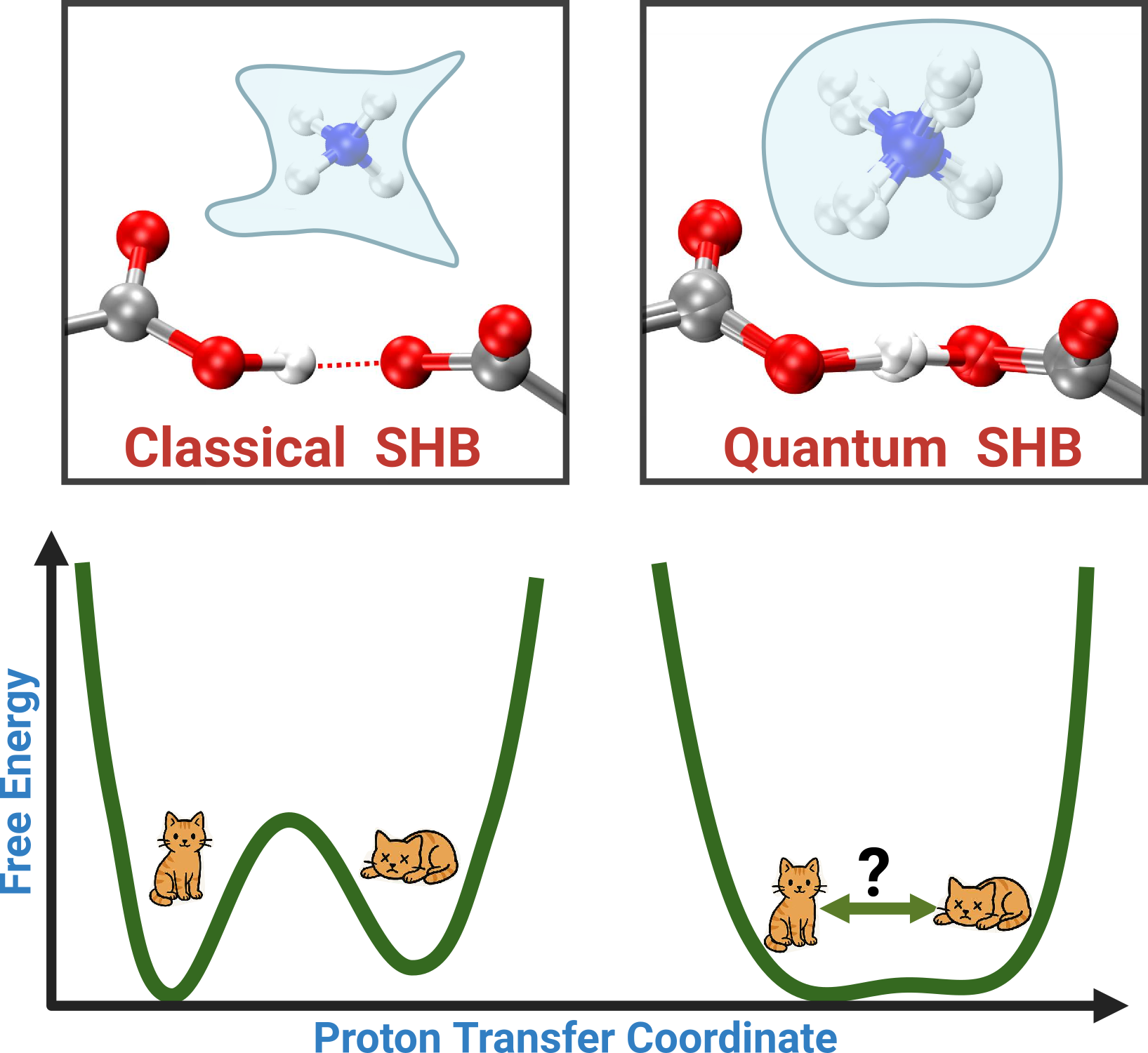}
\end{tocentry}

\begin{abstract}
Short hydrogen bonds, defined by donor–acceptor distances below 2.5 Å, represent a distinct regime in acid–base chemistry where conventional models of hydrogen bonding break down. In an organic crystal formed via a temperature-induced chemical transformation of L-glutamine, we previously identified a short hydrogen bond featuring a double-well potential indicative of activated proton transfer. Here, using path-integral \textit{ab initio} molecular dynamics, we show that nuclear quantum effects completely eliminate the classical barrier leading to a symmetrization of the proton along the hydrogen-bond. Classically, proton transfer is strongly coupled to the rocking motion of a neighboring ammonium ion; under quantum effects, this coupling is significantly reduced. Furthermore, examining the electronic structure through Wannier centers reveals a quantum-driven redistribution of bonding electrons, blurring the distinction between hydrogen bonding and covalency. Taken together, our findings indicate that nuclear quantum effects in this organic crystal create a regime where the donor and acceptor simultaneously act as acid and base.
\end{abstract}


Proton transfer (PT) is one of the most fundamental and pervasive processes in chemistry. It governs acid–base reactivity~\cite{cleland1994low,cox2009distance}, drives biochemical transformations and enzymatic catalysis~\cite{warshel2006electrostatic,wang2014quantum,fang2009mapping}, and plays a central role in solvation dynamics and energy transduction~\cite{ceriotti2016nuclear,marx2010aqueous,hassanali2013proton,siwick2008long,rini2003real}. At its core, PT is mediated by hydrogen bonds, typically viewed as asymmetric interactions between a donor and an acceptor, across which the proton hops by overcoming a well-defined energetic barrier~\cite{jeffrey1997an,intro_hbond_ishikita2014}.

This classical picture begins to break down when donor and acceptor atoms are brought into unusually close proximity~\cite{gilli2009nature,iogansen1999direct}, forming so-called short hydrogen bonds (SHBs) with donor–acceptor distances below $\sim$2.5\,\AA. In this regime, the potential energy surface flattens, the proton delocalizes, and nuclear quantum effects (NQEs) such as zero-point motion and tunneling become dominant~\cite{benoit1998tunnelling,dereka2021crossover,ceriotti2013nuclear,mouhat2023thermal}. Theoretical and spectroscopic studies show that this quantum crossover emerges when the PT barrier approaches the proton’s zero-point energy, and can be further modulated by external electric fields~\cite{dereka2021crossover, wang2014quantum,ceriotti2016nuclear,li2011quantum,cassone2020nuclear}. These have far-reaching consequences on both the physics and chemistry of a wide variety of systems. For example, proton symmetrization in high-pressure ice has been linked to O--O distances below 2.45\,\AA~\cite{benoit1998tunnelling}, engineered fluorescent proteins exhibit measurable SHB delocalization~\cite{wang2014quantum,oltrogge2015short}, and in water, NQEs modulate hydrogen bond strength and polarization~\cite{ceriotti2016nuclear,hassanali2013proton,agmon1995grotthuss}.

A lot of our understanding of SHBs comes from numerical atomistic simulations of liquid water solutions~\cite{giberti2014role,hassanali2013proton}, high-pressure phases of ice~\cite{benoit1998tunnelling,drechsel2014quantum}, inorganic crystal hydrates\cite{arandhara2024nuclear,saunders2021quantum,gurung2025quantum,buch2008hcl,hassanali2012fuzzy,kapil2022complete,rossi2016anharmonic}, in addition to small hydrated clusters\cite{li2024impact}. On the other hand, our understanding of SHBs in organic or biomolecular crystals, where hydrogen bonding can be shaped by a complex interplay of dense packing, vibrational disorder, as well as the local electrostatic potential, remain poorly understood. Although SHBs have been implicated in enzyme active sites~\cite{cleland1994low}, proton-relay chains~\cite{agmon1995grotthuss}, and fluorescent proteins~\cite{fang2009mapping}, their quantum mechanical behavior in crystalline environments remains largely uncharted. While spectroscopic studies on bifluoride salts~\cite{dereka2021crossover} and peptides~\cite{gorbitz1989hydrogen,sappati2016nuclear} have explored SHBs, few have investigated how the environmental effects in a solid-state crystalline system can be altered by nuclear quantum fluctuations.

In a recent study, we reported that thermal incubation of L-glutamine leads to the formation of a distinct organic crystal composed of pyroglutamic acid, pyroglutamate, and ammonium ions, which we termed L-pyro-amm~\cite{stephens2021short}. This structure features a chemically asymmetric SHB with a donor--acceptor distance of $\sim$2.5\,\AA, flanked by a centrally located ammonium ion. We found that the presence of SHBs in L-pyro-amm plays an important role in enhancing nonaromatic fluorescence, a phenomenon that is currently a very active area of research\cite{morzan2022non,tomalia2019non,tang2021nonconventional,wang2019reevaluating,lei2023evaluation,zhao2020aggregation,zheng2020accessing}. Specifically, our simulations revealed that the PT along the SHB as well as the vibrational distortions of the carbonyl bonds in close proximity\cite{miron2023carbonyl}, suppresses access to the conical intersections reducing nonradiative decay and increasing the fluorescence yield. This optical behavior is distinct from that of unmodified L-glutamine or Pyroglutamic acid crystals, neither of which display such fluorescence. Employing first principles simulations, we showed that the presence of the SHBs facilitates PT as a thermally activated process.

In this work, we revisit the acid-base chemistry we originally observed along the SHB using path-integral \textit{ab initio} molecular dynamics (PI-AIMD) simulations. We find that NQEs eliminate the classical PT barrier, producing an essentially symmetric potential with the proton localized in the middle of the SHB. While the classical PT is strongly coupled to the motions of the ammonium ion in the crystal due to directional hydrogen-bonds that it forms with the pyroglutamine moiety, NQEs reduce this coupling. Examining the Wannier centers (WCs) of the system, the electronic structure of the hydrogen-bond in the quantum simulations is characterized by features that give it both electrostatic and covalent character. This aspect blurs the distinction between the chemical moiety that acts as the proton donor and acceptor involved in the acid-base equilibrium that occurs between pyroglutamic acid and pyroglutamate. Taken together, these results suggest that SHBs embedded in soft crystalline frameworks could serve as platforms for quantum-responsive materials with tunable bonding, structure, and function.

\begin{figure*}
\centering
\includegraphics[clip=true, width=0.7\textwidth]{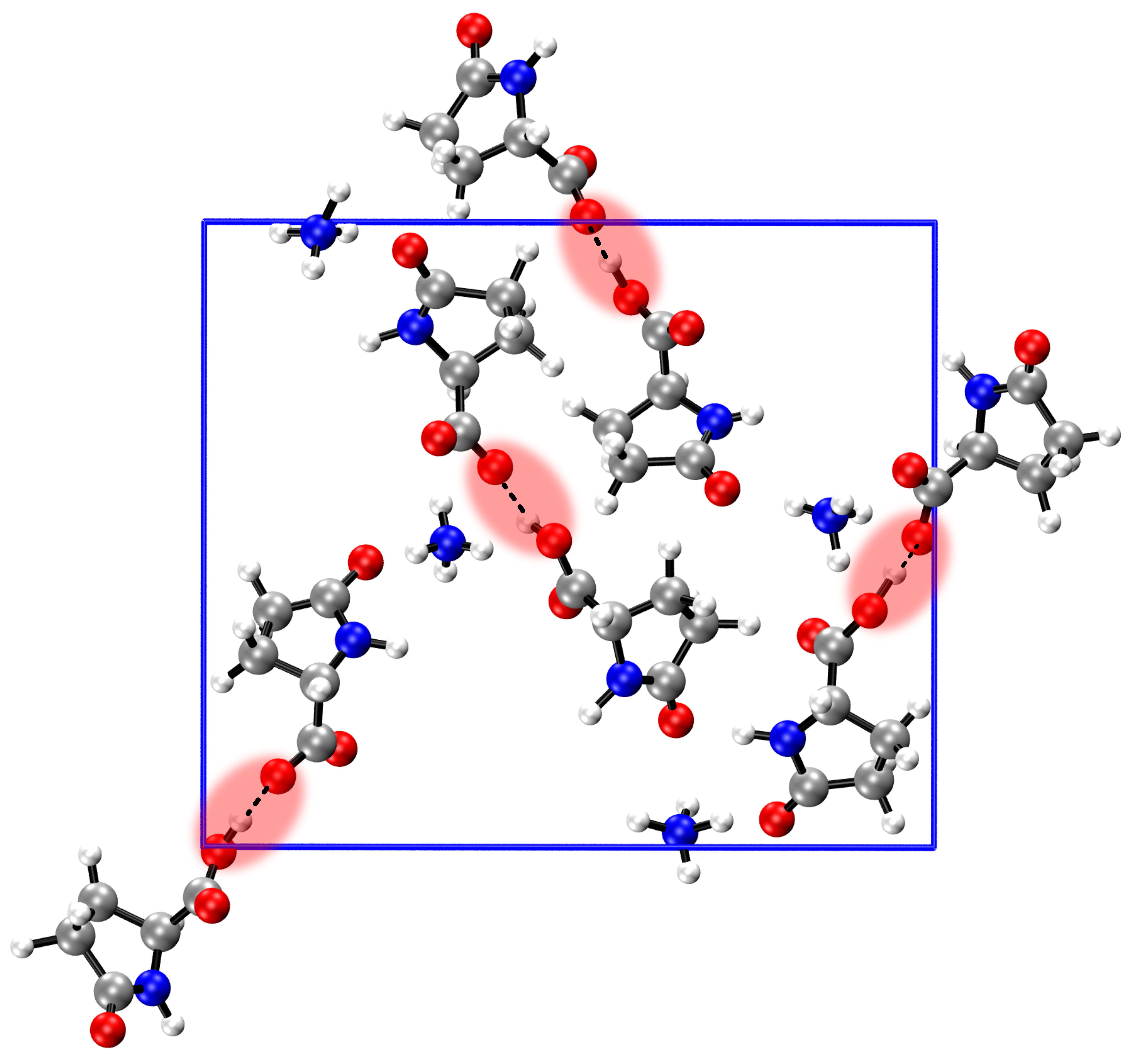}
\caption {Crystal structure of L-pyro-amm. SHBs are highlighted in the red shaded region. Each SHB is located in close proximity to an ammonium ion.}
\label{pyro-amm}
\end{figure*}

\section{Methods}
\label{methods}

To investigate the thermodynamic and structural properties of hydrogen bond networks in the studied systems, we performed \textit{ab initio} molecular dynamics (AIMD) simulations using the Quickstep module in the CP2K software package~\cite{vandevondele2005quickstep}. All simulations were carried out within the Born–Oppenheimer approximation. Electronic wavefunctions were expanded using a double-zeta valence polarized (DZVP) Gaussian basis set combined with a plane wave cutoff of 300~Ry for the electronic density. This combination balances computational efficiency and accuracy, and has been shown to reproduce key structural and vibrational properties in condensed-phase and hydrogen-bonded systems~\cite{vandevondele2007gaussian, marsalek2016ab, ceriotti2013nuclear}. To further support this choice, we performed benchmark single-point energy calculations on 10 representative snapshots in which proton transfer occurs along one of the short hydrogen bonds (SHBs). Energies were computed using DZVP and TZVP basis sets with electronic density cutoffs ranging from 300 to 450~Ry. The results, presented in the Supporting Information, show that all relative energy deviations remain below 0.015~eV, confirming the reliability of the DZVP/300~Ry setup for the present simulations.

Core electrons were modeled using Goedecker–Teter–Hutter pseudopotentials~\cite{goedecker1996separable}. Exchange correlation effects were treated using the Becke–Lee–Yang–Parr (BLYP) functional~\cite{lee1988development}, with Grimme D3(0) dispersion corrections~\cite{grimme2010consistent} to account for van der Waals interactions. All simulations were performed in the canonical (NVT) ensemble, with temperature (300 K) maintained using a velocity-rescaling thermostat~\cite{bussi2007canonical}. The integration timestep was 0.5~fs, and classical AIMD trajectories were run for 50~ps.

To incorporate NQEs, we employed path-integral \textit{ab initio} molecular dynamics (PI-AIMD)\cite{marx1996ab,tuckerman2002path}, wherein each nucleus is represented by a ring polymer of $P$ replicas (beads) connected by harmonic springs. This formalism enables quantum delocalization of light nuclei, particularly protons. To reduce the computational cost (which scales linearly with $P$), we used the PIGLET thermostat~\cite{ceriotti2012efficient}, which permits accurate sampling of nuclear quantum distributions with fewer beads. All PIMD simulations used $P = 6$, a setting validated in prior studies of hydrogen-bonded systems including liquid water~\cite{giberti2014role}. To assess convergence with respect to the number of beads, we also repeated key analyses with $P = 8$ and found that the resulting short hydrogen bond (SHB) free energy profiles remained unchanged (see SI Figure~2).
Production runs lasted approximately 10~ps, with trajectory data sampled after an initial equilibration period of 2 ps. For comparative purposes, the same computational approach was applied to L-glutamine (L-Glu) which consists of normal hydrogen bonds, as described in the SI.

To examine changes in electronic structure due to quantum fluctuations, we computed maximally localized WCs, which provide a real-space partitioning of the electronic density into chemically interpretable bonding and lone-pair components~\cite{marzari1997maximally,silvestrelli1999water}. The localization procedure was implemented using the built-in tools in CP2K. For each trajectory snapshot, we extracted the positions  of all WCs. In postprocessing, we analyzed the distributions of WCs assigned to the SHB-forming oxygen atoms, distinguishing between those involved in O--H bonding and those corresponding to lone pairs. These distributions were then compared for configurations coming from both the classical and quantum simulations.


We performed classical (AIMD) and quantum (PI-AIMD) simulations to investigate NQEs in glutamine-derived crystal structures. The first, L-glutamine, crystallizes in the orthorhombic P2$_1$2$_1$2$_1$ space group, forming an extended network in which each glutamine molecule participates in approximately five N--H$\cdots$O hydrogen bonds. These donor--acceptor distances range from 2.7 to 2.9\,\AA~\cite{cochran1952crystal}, consistent with conventional asymmetric hydrogen bonding (see SI Figure 1a).

\begin{figure*}
\centering
\includegraphics[width=\textwidth]{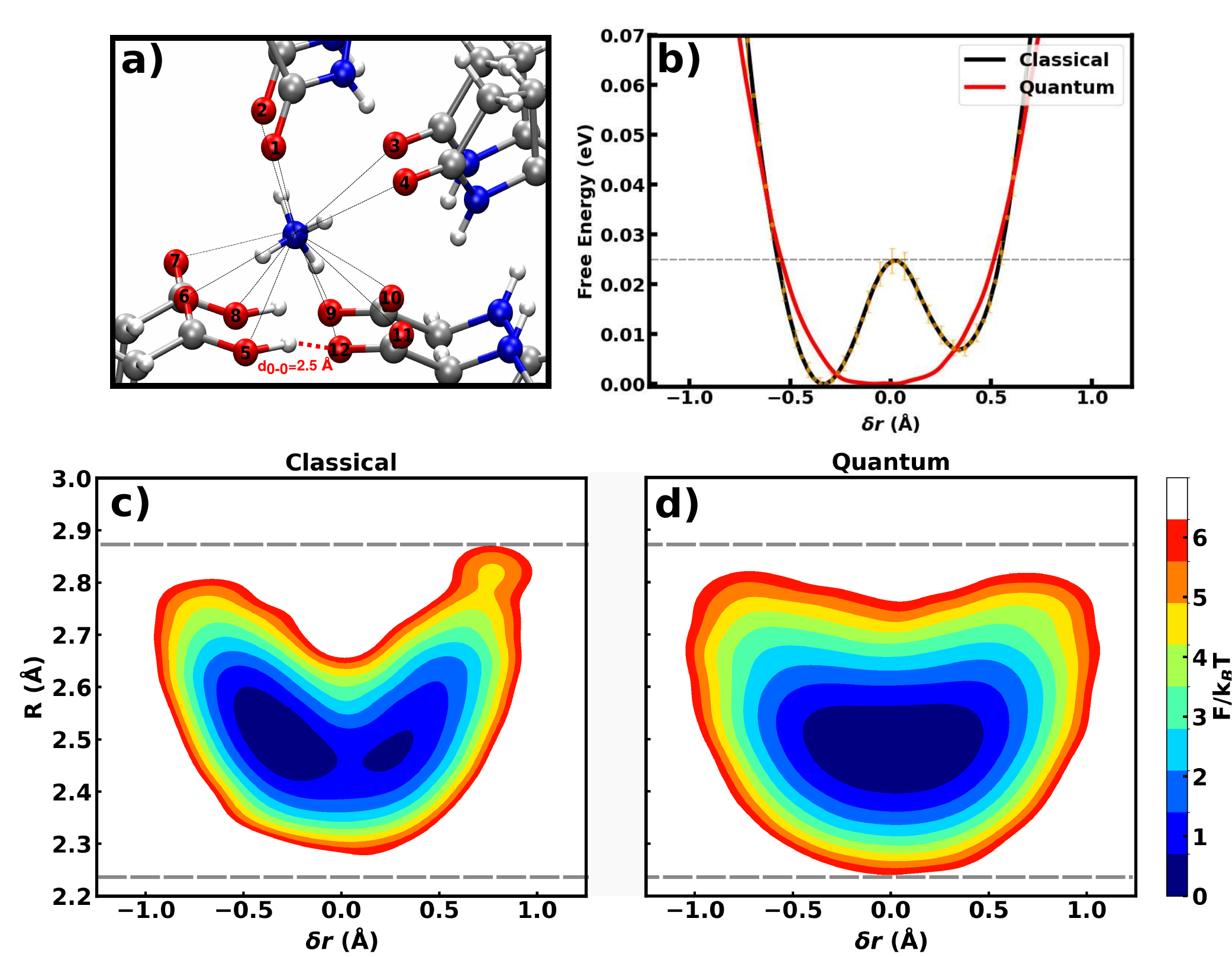}
\caption{
(a) A zoomed snapshot of the L-pyro-amm structure highlighting the SHB (dashed circle) formed between two carboxylate oxygen atoms. The surrounding ammonium ion engages in hydrogen bonding interactions with twelve oxygen atoms located within a 5\,\AA\ radius, including the SHB oxygens. These interactions form a complex local environment modulating SHB dynamics.
(b) One-dimensional free energy profiles along the PT coordinate from classical (black) and quantum (red) simulations. The error bars of the classical free energy profile were estimated using block analysis with a block size of 50 fs. The horizontal dashed line refers to the $k_{B}T$ value at 300 K.
(c,d) Joint distributions of the PT coordinate $\delta r$ and the heavy atom compression coordinate $R = d_{\mathrm{X{-}X'}}$ under classical and quantum conditions, respectively. Dashed horizontal lines in (c) indicate the O--O distance range sampled in classical (upper dashed line) and quantum simulations (lower dashed line).
}
\label{fep}
\end{figure*}

Upon incubation at 60\,\textdegree C, L-glutamine undergoes a cyclization reaction, yielding a distinct organic crystal composed of pyroglutamic acid, pyroglutamate, and ammonium ions (details are published in our previous work~\cite{stephens2021short}). This structure, L-pyro-amm, retains several conventional hydrogen bonds but also features a chemically distinct SHB between two carboxylate oxygen atoms, with a donor--acceptor distance of approximately 2.5\,\AA\ (see Figure~\ref{pyro-amm} and Figure~\ref{fep}a). Crucially, this SHB is embedded within a local electrostatic field shaped by a neighboring ammonium ion, which donates hydrogen bonds to twelve nearby oxygen atoms—forming a dynamic, multidirectional interaction network. In previous work, we demonstrated that this SHB is correlated with non-aromatic fluorescence in the crystal, suggesting a possible interplay between hydrogen bonding and optoelectronic properties.~\cite{stephens2021short}

To quantify the behavior of the SHB, we define a PT coordinate as $ \delta r = d_{\mathrm{X{-}H}} - d_{\mathrm{H{-}X'}}$, where X and $X'$ are the two oxygen atoms forming the SHB. A negative value of $\delta r$ indicates that the proton is localized on the donor; $\delta r = 0$ corresponds to a perfectly shared proton, and positive values correspond to transfer to the acceptor. The free energy profiles along this coordinate are shown in Figure~\ref{fep}b. Classically, the SHB is characterized by a low-barrier, asymmetric double-well potential, with the proton biased toward one side. The barrier height is approximately 30\,meV, significantly below the zero-point energy (ZPE) of an O--H stretch—suggesting that quantum fluctuations may alter this landscape.

Indeed, when NQEs are included via path-integral dynamics, the double-well is replaced by a nearly symmetric, single-well distribution. The proton is shared between donor and acceptor, reflecting a quantum-delocalized state that erases the barrier typically associated with classical acid–base transfer. This transition is consistent with proton symmetrization previously observed in high-pressure ice~\cite{benoit1998tunnelling,cherubini2024quantum}, but here it occurs within a chemically complex organic environment under room temperature conditions. For comparison, we also computed the PT free energy profile for L-glutamine (L-glu), which contains standard hydrogen bonds. Unlike the SHB system in L-pyro-amm, L-glu exhibits a single-well potential, indicating no PT. Full details are provided in the Supporting Information (see SI Figure 3).

\begin{figure}[h]
\centering
\includegraphics[width=\textwidth]{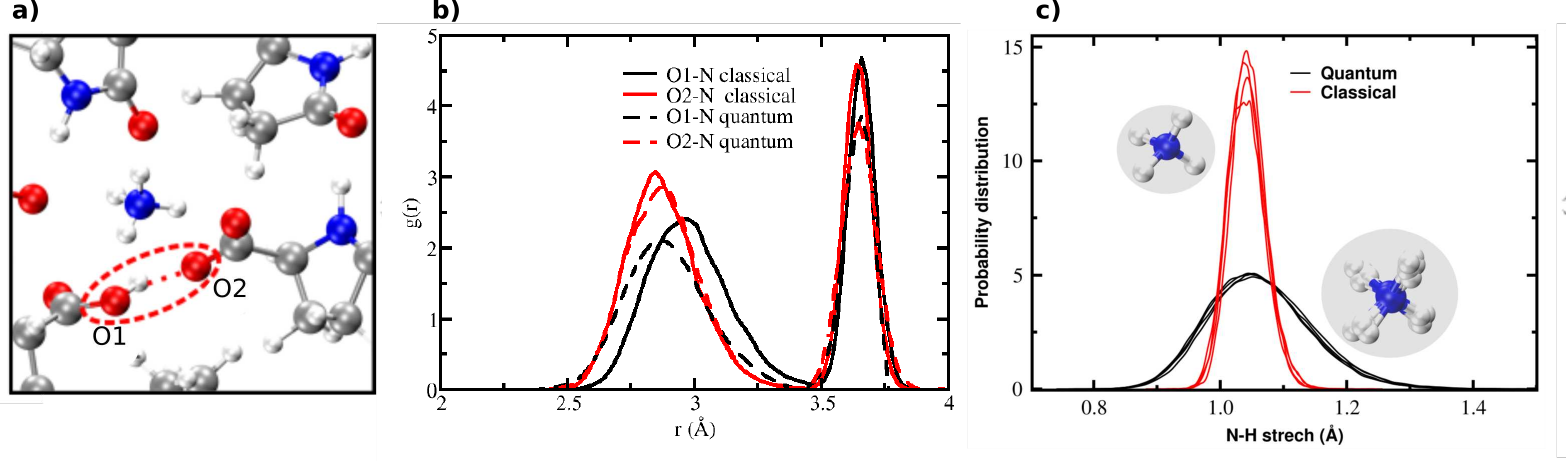}
\caption{(a) A zoomed crystal snapshot showing SHB and ammonium environment. (b) RDFs of SHB oxygen atoms with surrounding nitrogen atoms in classical (solid) and quantum (dashed) simulations. (c) Distribution of N--H bond lengths in the ammonium ion from classical (red) and quantum (black) simulations. The quantum distribution is significantly broadened (illustrated by a representative structure snapshot for clarity), reflecting enhanced proton delocalization and the effective ``expansion'' of the ionic radius. }
\label{rdf}
\end{figure}

To probe the origin of the classical asymmetry of the SHB in L-pyro-amm, we analyzed the surrounding chemical environment by computing radial distribution functions (RDFs) between each SHB oxygen atom (Figure~\ref{rdf}a) and all nitrogen atoms in the crystal. As shown in Figure~\ref{rdf}b, the ammonium ion interacts more strongly with one of the oxygen atoms (O2), shifting its RDF peak to shorter distances relative to the other (O1). This asymmetry biases the proton toward a slightly higher probability of localization on O2. Under quantum conditions, these RDFs become more similar—indicating partial spatial delocalization of the ammonium ion, as evidenced by broader N--H bond distributions (Figure~\ref{rdf}c), and a reduction in environmental asymmetry.


Besides the environmental asymmetry, we also examined how the PT correlates with the distance between the two SHB oxygen atoms. Specifically, we analyzed the joint distribution of the PT coordinate ($\delta r$) and the heavy atom compression coordinate $R = d_{\mathrm{X{-}X'}}$, shown in Figures~\ref{fep}c and \ref{fep}d. In classical simulations, PT is tightly correlated with O--O distance contraction, with the bond compressing toward $\sim$2.5\,\AA\ at the transition state. When NQEs are included, the distribution broadens, and the average O--O distance shifts slightly lower, consistent with a modest strengthening of the hydrogen bond. The correlation between O--O compression and proton position thus weakens, indicating that quantum fluctuations enable proton delocalization without the structural distortion of the hydrogen bond. This softening of coupling between heavy atom motion and proton sharing is consistent with previous findings~\cite{ceriotti2013nuclear,marx2006proton}.

While the structural asymmetry of the SHB diminishes under quantum conditions, one question that remains is how exactly the PT is coupled to the environment, in this particular case, the presence of the positively charged ammonium ion. To do this, we next examined how quantum fluctuations modify the structure of the ammonium ion itself. The reduction in local environmental asymmetry observed under quantum conditions raises an important mechanistic question: how do NQEs alter the ammonium ion’s interactions with its surroundings? To explore this, we analyzed the distribution of N--H bond lengths in the ammonium ion from classical and quantum simulations. As shown in Figure~\ref{rdf}c, quantum fluctuations significantly broaden these bond length distributions, consistent with increased proton delocalization around the nitrogen center. NQEs thus make the ammonium ion slightly larger which reduce the difference of proximity of the ion to the oxygen atoms forming the SHB.

Does this spatial broadening reduce the directionality of hydrogen bonding between the ammonium ion and neighboring SHB oxygen atoms? To answer this, we analyzed how PT correlates with the internal motion of the ammonium ion. Specifically, we computed the distances between the ammonium nitrogen and all the surrounding oxygen atoms ($d_i$), and examined several distances defined as $d_{i,j} = d_i - d_j$, which capture internal rocking or compression modes of the ion relative to the SHB.

Figures~\ref{corr}a and \ref{corr}b show joint probability distributions between the proton transfer coordinate ($\delta r$) and one representative ammonium rocking mode ($d_{6,10}$), comparing classical and quantum conditions. In classical simulations, $\delta r $ is tightly coupled to this mode, with PT events coinciding with directional shifts of the ammonium ion. This behavior resembles a pre-solvation mechanism previously observed for the PT mechanism in bulk water. \cite{agmon1995grotthuss,berkelbach2009concerted,day2000mechanism,kornyshev2003kinetics,swanson2007proton,marx2006proton,kabbe2017proton}. As the ammonium ion moves toward one SHB oxygen, it transiently stabilizes the conjugate base, facilitating proton hopping. In contrast, quantum simulations reveal a broader, more diffuse distribution, with significantly reduced correlation between $\delta r$ and ammonium motion.


To systematically assess the extent of this coupling, we computed Pearson correlation coefficients between the PT coordinate $\delta r$ and all pairwise ammonium–oxygen distance differences ($d_{i,j}$), shown in Figures~\ref{corr}c (classical) and \ref{corr}d (quantum). The Pearson correlation coefficient between $\delta r$ and a given distance difference $d_{i,j}$ is defined as:
\begin{equation}
\rho_{\delta r,\,d_{i,j}} = \frac{\mathrm{Cov}(\delta r,\, d_{i,j})}{\sigma_{r} \, \sigma_{d_{i,j}}},
\end{equation}
where $\mathrm{Cov}(\delta r,\, d_{i,j})$ is the covariance between the PT coordinate $\delta r$ and the distance-differences--$d_{i,j}$ defined earlier, and $\sigma_{r}$ and $\sigma_{d_{i,j}}$ are their respective standard deviations. In the classical simulations, several modes show strong correlation with the PT coordinate $\delta r$, with coefficients as high as 0.68 (e.g., $d_{6,10}$ and $d_{2,8}$ in the SHB-aligned direction). Under quantum conditions, these values drop substantially, with most correlations falling below 0.35 and many near or below 0.2. This global reduction reflects a scenario in which quantum fluctuations flatten the proton potential and reduce its sensitivity to the fluctuations of the ionic environment.

\begin{figure*}
\centering
\includegraphics[width=\textwidth]{fig4.pdf}
\caption{
(a,b) Joint probability distributions between the PT coordinate ($\delta r$) and the ammonium rocking mode defined by the distance difference $d_{6,10} = d_6 - d_{10}$ for classical and quantum simulations, respectively. Classically, strong correlation indicates pre-solvating behavior by the ammonium ion; quantum effects disrupt this coupling. Other ammonium rocking modes with high Pearson correlation coefficients exhibit similar coupling behavior with the PT coordinate. Free energy surfaces for these additional modes are provided in the Supporting Information. (c,d) Pearson correlation matrices between $\delta r$ and all pairwise ammonium-oxygen distances $d_{i,j}$ for classical and quantum simulations, respectively. Stronger correlations appear along the oxygen atoms aligned with the SHB axis (e.g., $d_{6,10}$, $d_{2,8}$). Quantum fluctuations globally reduce these couplings.
}
\label{corr}
\end{figure*}

The barrierless proton sharing observed in L-pyro-amm under quantum conditions should also be reflected in the underlying electronic structure of the SHB. As the proton becomes more delocalized between donor and acceptor atoms, the geometric symmetrization of the SHB is accompanied by a redistribution of electron density. To examine this, we analyzed the spatial distributions of maximally localized WCs associated with the SHB-forming oxygen atoms. These WCs provide a real-space representation of bonding electrons and enable a detailed picture of how the electronic structure responds to quantum fluctuations.

In a typical hydrogen-bonded system—such as water—each oxygen atom is assigned four WCs: two associated with lone pairs and two with covalent bonds to protons. For standard O--H$\cdots$O hydrogen bonds longer than 2.8~\AA{}, the WC distributions are well-separated: lone pairs cluster near the oxygen nucleus, while the bonded electrons extend toward neighboring atoms as discussed in previous findings~\cite{silvestrelli1999water}.

\begin{figure*}
\centering
\includegraphics[width=0.8\textwidth]{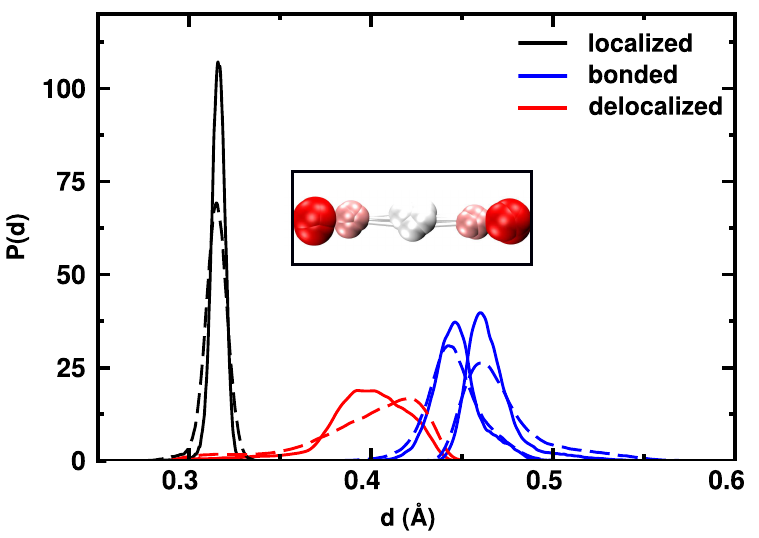}
\caption{
Distribution of distances between WCs and the oxygen atom in the SHB of L-pyro-amm, from classical (solid lines) and quantum (dashed lines) simulations. Each oxygen hosts four WCs: two associated with covalent bonds (C--O, O--H) and two lone pairs. The bonded WC associated with the O--H interaction becomes delocalized and shifts outward under quantum conditions, reflecting elongation and partial covalent character of the SHB. The inset visualizes this environment in the quantum simulation: red spheres are oxygen atoms, white are protons, and pink spheres are WC positions. For quantum case, the WC distributions are obtained by averaging over all six beads.
}

\label{fig5}
\end{figure*}

In contrast, the SHB in L-pyro-amm exhibits markedly different WC behavior. Figure~\ref{fig5} shows the distance distributions of the four WCs assigned to one SHB oxygen atom. In classical simulations (solid lines), the WC associated with the O--H bond peaks sharply near 0.4\,\AA, while lone pair WCs remain localized near 0.3\,\AA. Upon inclusion of NQE (dashed lines), the O--H bonded WC broadens and shifts outward, reflecting elongation and partial delocalization of the bond. The WC distributions overlap more significantly, indicating increased electronic polarization and a departure from conventional hydrogen bonding toward partial covalency.

This reorganization of bonding electrons is consistent with the delocalized proton distribution observed in the quantum simulations and highlights the coupling between nuclear and electronic degrees of freedom. Notably, this WC polarization is unique to the SHB: in other hydrogen bonds in the crystal, bonded and non-bonded WCs remain well-separated regardless of quantum treatment (see SI Figures 4 ).

For comparison, we also analyzed WC behavior and PT energetics in L-glutamine, which contains only conventional hydrogen bonds. These results, shown in the Supporting Information (SI Figures 5), confirm the absence of significant proton delocalization or WC overlap in L-glutamine, even under quantum conditions. This supports the conclusion that quantum-driven electronic rehybridization is not a general feature of hydrogen bonds but arises under specific geometrical and electrostatic environments, as exemplified by the SHB in L-pyro-amm.

\section{Conclusion}
\label{conclusion}

SHBs represent a class of interactions where new chemistry can emerge. Specifically in these hydrogen bonds, classical
models of PT do not adequately describe the physics and chemistry underlying these interactions. In this work, we investigated a glutamine-derived organic crystal containing a chemically asymmetric SHB coupled to a nearby ammonium ion. Using path-integral \textit{ab initio} molecular dynamics simulations, we showed that NQEs eliminate the classical double-well proton potential, resulting in a nearly symmetric, barrierless distribution. All these aspects, lead to the emergence of rather exotic situation, induced by NQEs, where a chemical system behaves simultaneously as an acid and a base.

Beyond this central result, our study highlights a subtle interplay between quantum fluctuations and the local ionic environment. Under quantum conditions, the ammonium ion, which classically stabilizes the SHB via directional hydrogen bonding and low-frequency rocking modes, exhibits structural broadening. This broadening reduces the directionality of its electrostatic interactions and leads to a more symmetric chemical environment around the SHB, which in turn facilitates the symmetrization of the proton.

Our analysis of maximally localized WCs further reveals the important role of quantum effects in tuning the underlying electronic structure of SHBs. The redistribution of bonding electrons—marked by increased overlap between lone pair and covalent WCs—illustrates how quantum nuclear fluctuations induce partial covalent character in what classically behaves as a hydrogen bond. This provides a nice model system to demonstrate and explore how NQEs couple with the underlying electronic structure.

Together, these findings identify a distinct behavior of SHBs leading to acid–base ambiguity in organic crystals in which SHBs are not static entities but dynamic, quantum-responsive bonding motifs shaped by their chemical environment. To our knowledge, this is the first atomistic demonstration of quantum-induced, barrierless proton delocalization in an organic SHB modulated by a counterion. This mechanism—quantum–environment decoupling—may be more widespread and functionally relevant than previously appreciated, particularly in biological systems, hydrogen-bonded molecular crystals, and proton-coupled electronic materials.

Looking forward, this work suggests that SHBs can be engineered as tunable elements in materials design. By modulating local degrees of freedom—such as ionic composition, lattice flexibility, or electronic polarizability, it may be possible to program SHBs with tailored structure and reactivity for applications in sensing and opto-electronics. Within the context of the phenomenon of non-aromatic fluorescence, the presence of an SHB where the proton is equally shared and with an electronic structure that has partial covalent character, makes the hydrogen bonds stronger and enhances electron delocalization. Therefore NQEs in principle may enhance the possibility of reducing non-radiative decay. A detailed TDDFT-based analysis exploring these fluorescence-related effects is currently underway and will be presented in a forthcoming study.

Finally, this study opens several avenues for future exploration: How general is this mechanism across different chemical classes? Can similar effects be realized in biomolecular systems or hybrid materials? Do quantum fluctuations influence cooperative PT networks or coupled electron–proton transport phenomena? In a recent study reported by some of us, the amino-acid Cysteine, crystallized in light (H2O) and heavy (D2O) waters yields strikingly different crystal structures and subsequently, optical properties. The role of quantum effects in stabilizing certain hydrogen-bond networks in these types of organic crystals would be and interesting area to explore given the findings of this work. Addressing these questions will require tighter integration of quantum simulation techniques with experimental approaches capable of resolving proton and electron dynamics—an effort that this work helps to motivate and benchmark.

\section*{Data Availability}

The classical and quantum molecular dynamics simulations trajectories, along with the analysis scripts used in this study, are available from the corresponding author upon reasonable request.

\section*{Conflict of Interest}

The authors declare no competing interest.

\begin{acknowledgement}
A.H. acknowledges funding from the European Research Council (ERC) under the European Union’s Horizon 2020 research and innovation programme (grant agreement No. 101043272 – HyBOP). The views and opinions expressed are those of the authors only and do not necessarily reflect those of the European Union or the European Research Council Executive Agency. Neither the European Union nor the granting authority can be held responsible for them.
\end{acknowledgement}

\begin{suppinfo}
The Supporting Information provides additional simulation details, including crystal structures of L-glutamine and L-pyro-amm, convergence analysis of the proton free energy with respect to the number of beads used in path-integral molecular dynamics (PIMD) simulations, free energy profiles for L-glutamine. extended Wannier center analysis and structural correlations between proton transfer and ammonium vibrational modes, Figures S1–S6 support the findings discussed in the main text.
\end{suppinfo}


\bibliography{main}
\clearpage
\section{Supporting Information}













\section*{1. Crystal Structures of L-glutamine and L-pyro-amm}

\begin{figure}[h]
\centering
\includegraphics[width=\textwidth]{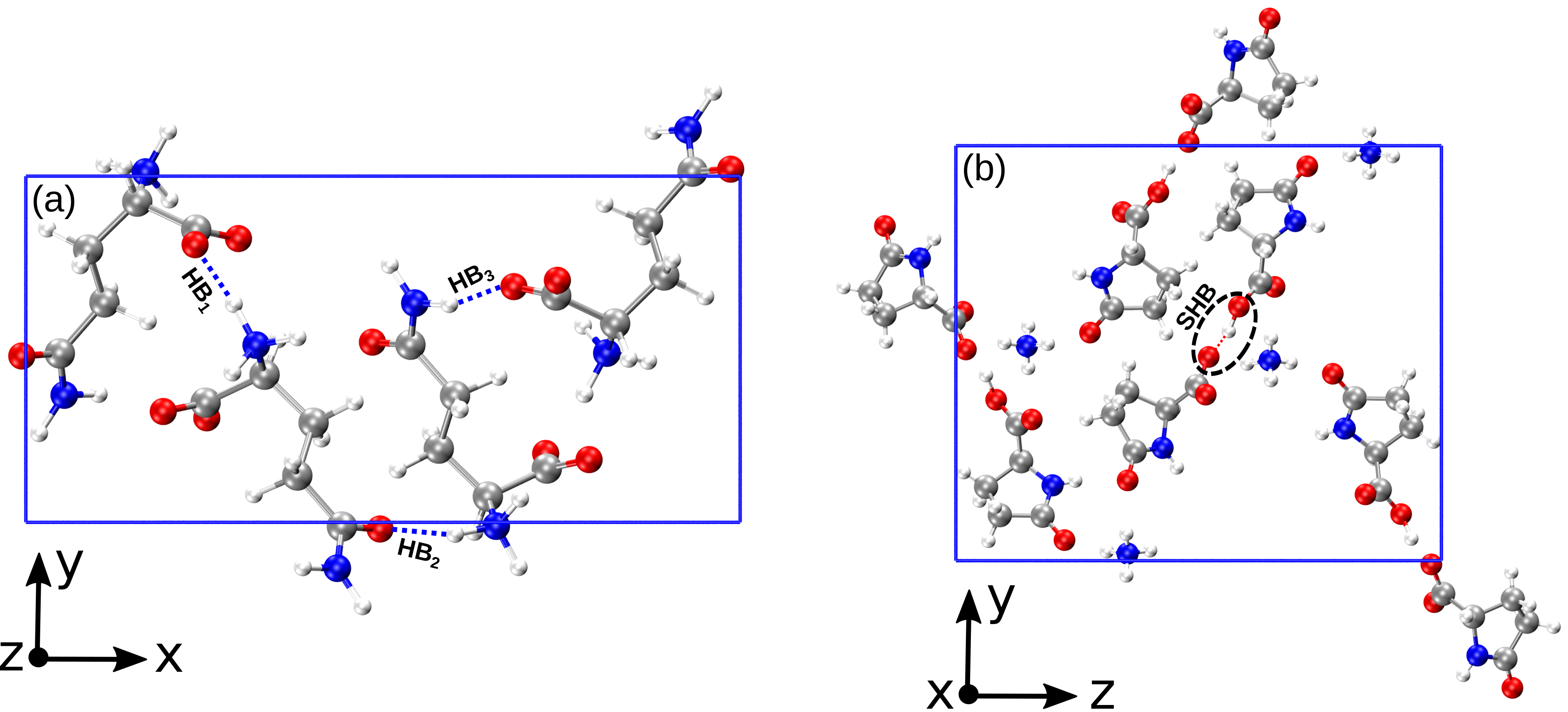}
\caption{(a) Crystal structure of L-glutamine with hydrogen bonds HB1--HB3 (2.7–2.9\,\AA). (b) Crystal structure of L-pyro-amm highlighting the SHB and nearby ammonium ion.}
\label{fig:crystals}
\end{figure}

L-glutamine crystallizes in the orthorhombic P2$_1$2$_1$2$_1$ space group with four molecules in the unit cell\cite{cochran1952crystal}. These molecules are held together by an extended network of hydrogen bonds. Figure~\ref{fig:crystals}(a) presents the unit cell highlighting three representative hydrogen bonds (HB1--HB3). These hydrogen bonds span distances between 2.7 and 2.9,\AA, which is typical for standard N--H$\cdots$O hydrogen bonds and remain well-localized even under quantum conditions. The nature and strength of these hydrogen bonds form a baseline for understanding deviations observed in the thermally derived phase.

Upon thermal incubation at 60,\textdegree C, L-glutamine undergoes cyclization, forming a new crystal composed of pyroglutamic acid, pyroglutamate, and ammonium ions\cite{stephens2021short}. The resulting L-pyro-amm structure also adopts the orthorhombic P2$_1$2$_1$2$_1$ symmetry. Figure~\ref{fig:crystals}(b) shows this new crystal configuration. Unlike L-glutamine, L-pyro-amm contains a chemically distinct short hydrogen bond (SHB) of approximately 2.5,\AA\ formed between two carboxylate oxygen atoms. This SHB is embedded within a highly asymmetric local environment influenced by a nearby ammonium ion, which interacts with twelve different oxygen atoms, creating a complex hydrogen bonding network.

\section*{2. Convergence of Path Integral Simulations}

\begin{figure}[h]
\centering
\includegraphics[width=0.65\textwidth]{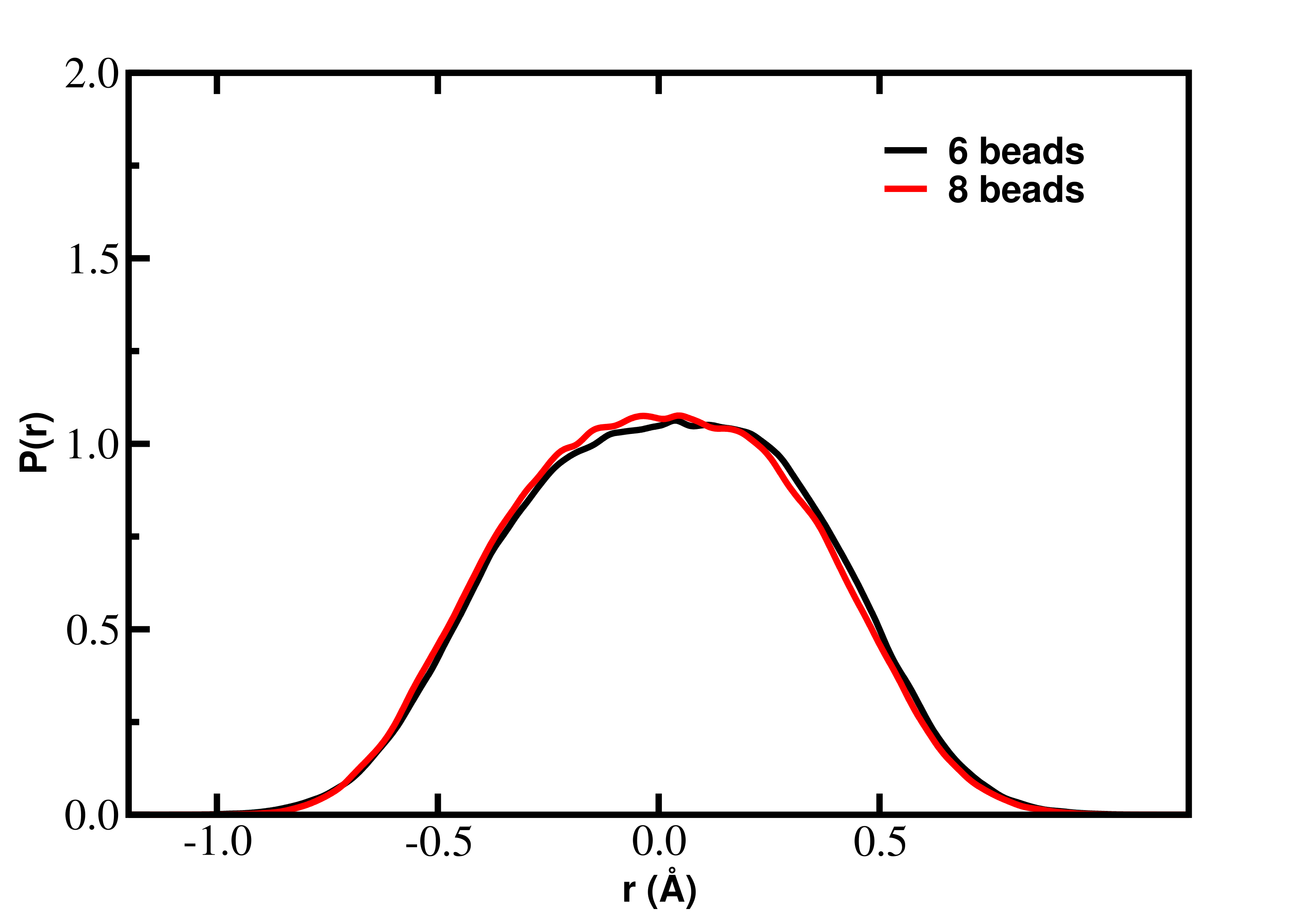}
\caption{Proton transfer coordinate distributions for SHB with 6 and 8 beads in PIGLET simulations.}
\label{fig:convergence}
\end{figure}

Ensuring convergence with respect to the number of beads used in path integral simulations is essential for quantitative reliability. We compared proton transfer coordinate distributions obtained using 6 and 8 beads within the PIGLET thermostat. Figure~\ref{fig:convergence} confirms that the 6-bead simulation captures the key structural features of proton delocalization, with negligible deviations from the more computationally expensive 8-bead run. This validates our use of 6 beads in all production simulations.\\

\begin{figure}[h]
\centering
\includegraphics[width=\textwidth]{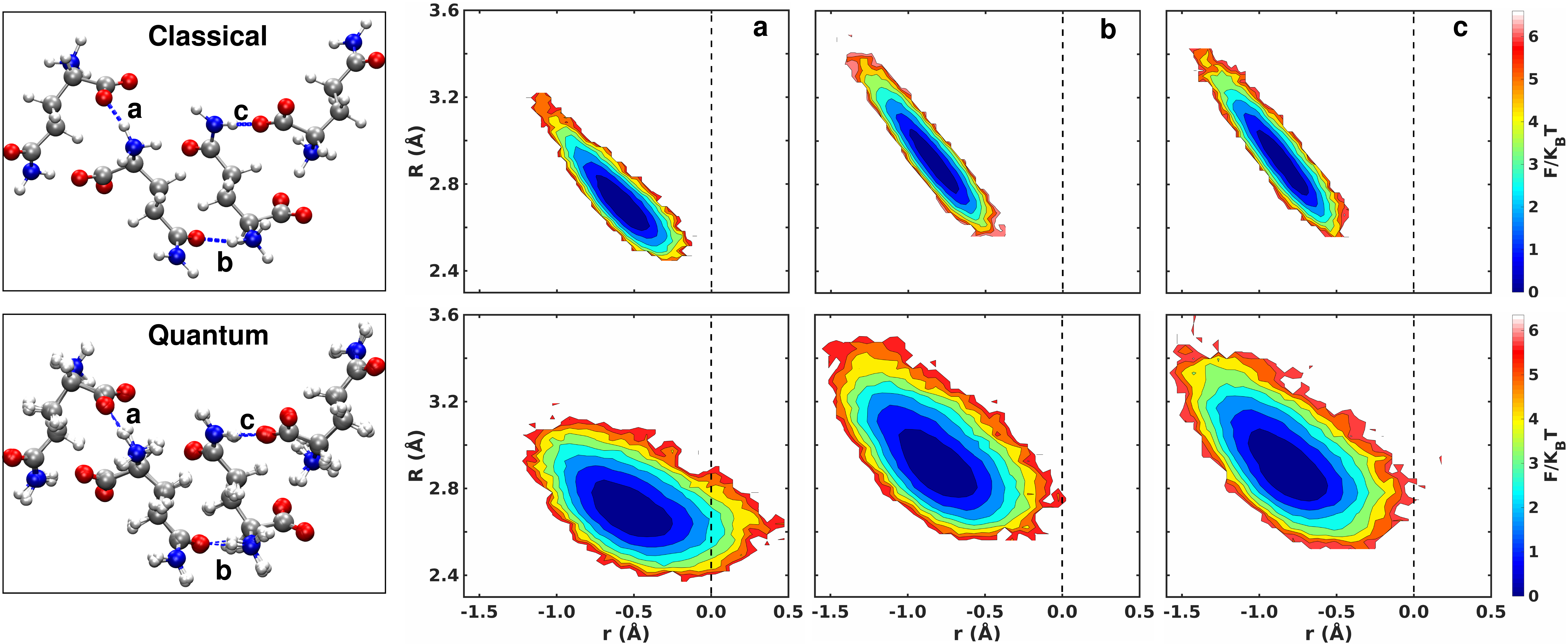}
\caption{Free energy profiles of proton transfer coordinate for SHB in L-glutamine along its different hydrogen bonds, under classical and quantum simulations. No barrier suppression observed.}
\label{fig:glu-ptc}
\end{figure}

\section*{3. Proton Transfer Profiles in L-glutamine}

To contextualize the unique behavior of the SHB in L-pyro-amm, we also examined proton transfer potentials in L-glutamine. As shown in Figure~\ref{fig:glu-ptc}, even the shortest hydrogen bonds in L-glutamine remain single-welled with no barrier suppression, consistent with a conventional, localized proton. This further supports the conclusion that SHB symmetrization observed in L-pyro-amm is not merely a consequence of reduced bond length, but results from a combination of geometric constraint and local electrostatic modulation by the ammonium ion.

\begin{figure}[h]
\centering
\includegraphics[width=0.65\textwidth]{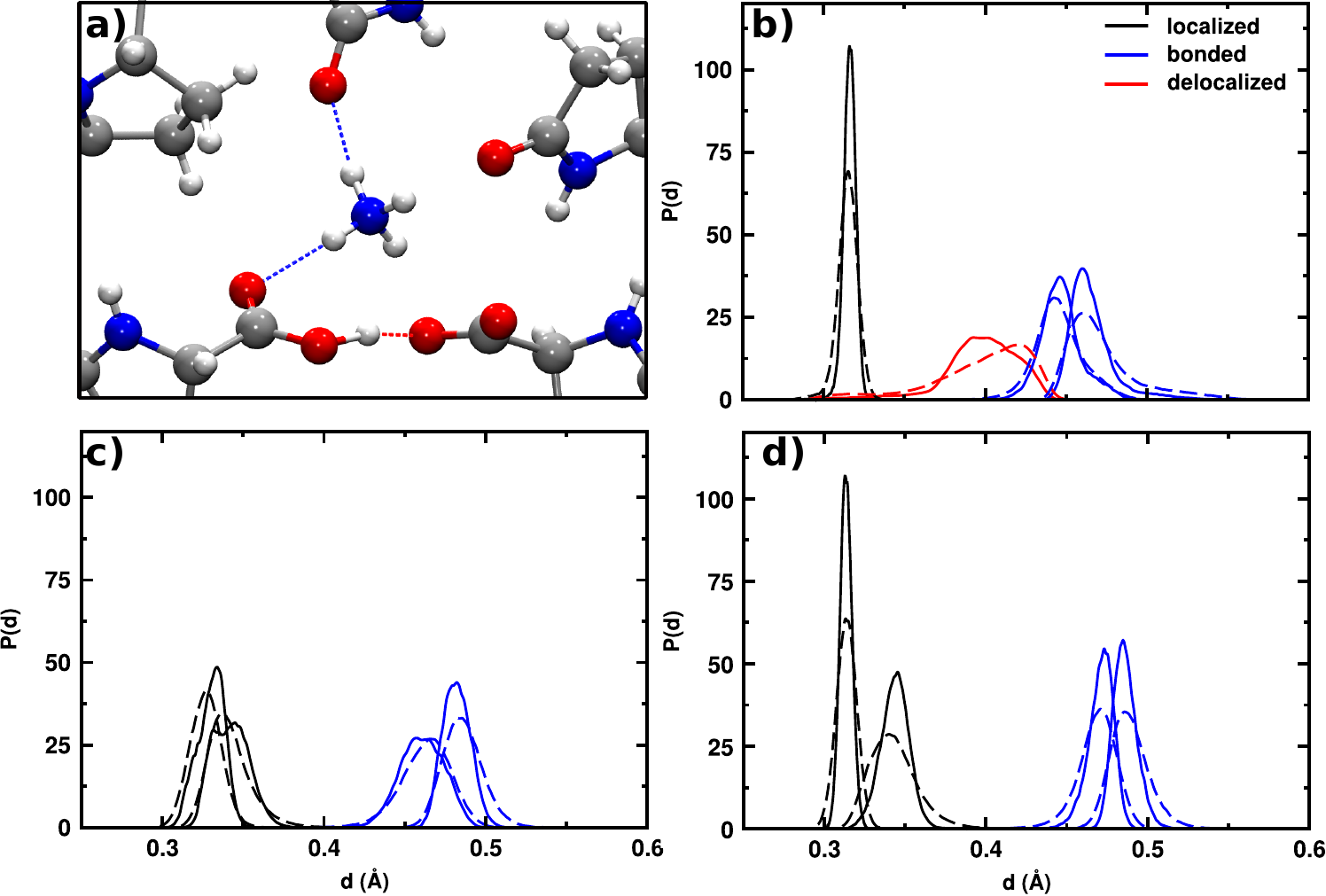}
\caption{
\textbf{(a)} Zoomed-in structural snapshot of L-pyro-amm showing representative hydrogen bonds formed between the ammonium ion and surrounding oxygen atoms (blue dashed lines), as well as the short hydrogen bond (SHB) highlighted with a red dashed line.
\textbf{(b)} Classical (solid lines) and quantum (dashed lines) distributions of Wannier centers projected along the O--H bond axis for the SHB oxygen.
\textbf{(c, d)} Wannier center distributions for the normal hydrogen bonds involving the ammonium ion, illustrating differences between classical and quantum conditions.
}

\label{fig:wan-shb}
\end{figure}

\begin{figure}[h]
\centering
\includegraphics[width=0.65\textwidth]{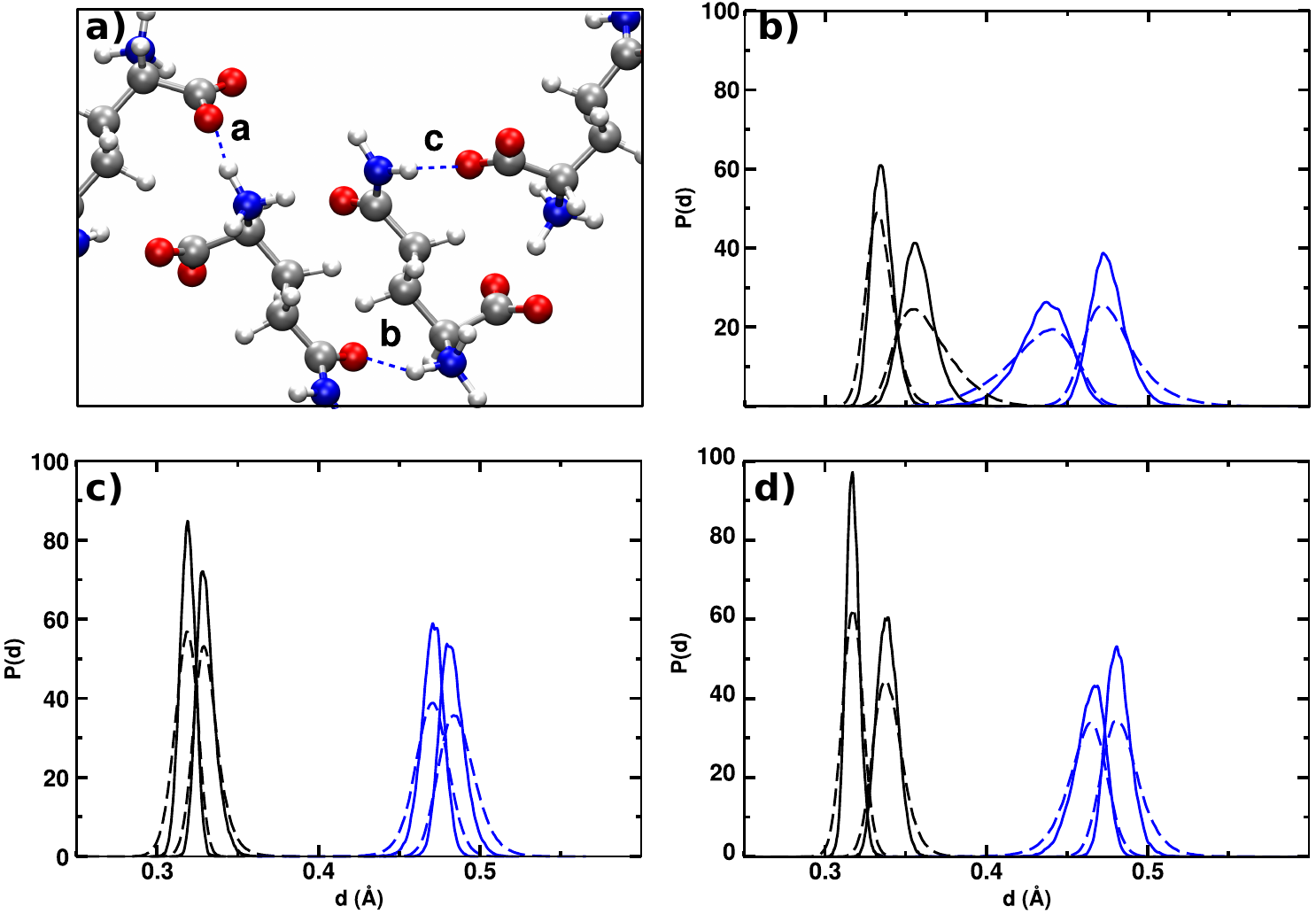}
\caption{
\textbf{(a)} Zoomed-in structural snapshot of L-glu showing its hydrogen bonds (a, b, and c) formed between terminal groups and between terminal and side-chain groups.
\textbf{(b)} Distribution of Wannier centers for the terminal hydrogen bond, showing strong polarization compared to the hydrogen bonds involving side-chain groups (b and c), as shown in panels (c) and (d), respectively.
}

\label{fig:wan-lglu}
\end{figure}

\section*{4. Wannier Center Distributions: SHBs vs. Normal Hydrogrn Bonds}

To probe electronic structure changes induced by quantum fluctuations, we analyzed the distributions of maximally localized Wannier centers (WCs) associated with oxygen atoms in hydrogen bonds. Figure~\ref{fig:wan-shb} shows the WC distributions for the SHB oxygen atom in L-pyro-amm (a same showin the main text), as well as for the hydrogen bonds formed between the ammonium ion and surrounding oxygen atoms. Classical simulations exhibit well-separated peaks corresponding to lone pairs and bonded electron pairs. In contrast, the quantum distributions are broadened, with partial overlap between lone pair and bonded WCs—indicating a delocalization of electron density and a transition from hydrogen bonding toward partial covalency.

For comparison, Figure~\ref{fig:wan-lglu} presents WC distributions for the hydrogen bonds in L-glutamine. Even under quantum conditions, these hydrogen bonds maintain a clear separation between lone pair and bonded WCs, demonstrating that the behavior observed for the SHB in L-pyro-amm is not universal, but rather dependent on specific geometric and electrostatic environments.

\section*{5. Structural Correlation Between Proton Transfer and Ion Vibrational Modes}

\begin{figure}[h]
\centering
\includegraphics[width=0.65\textwidth]{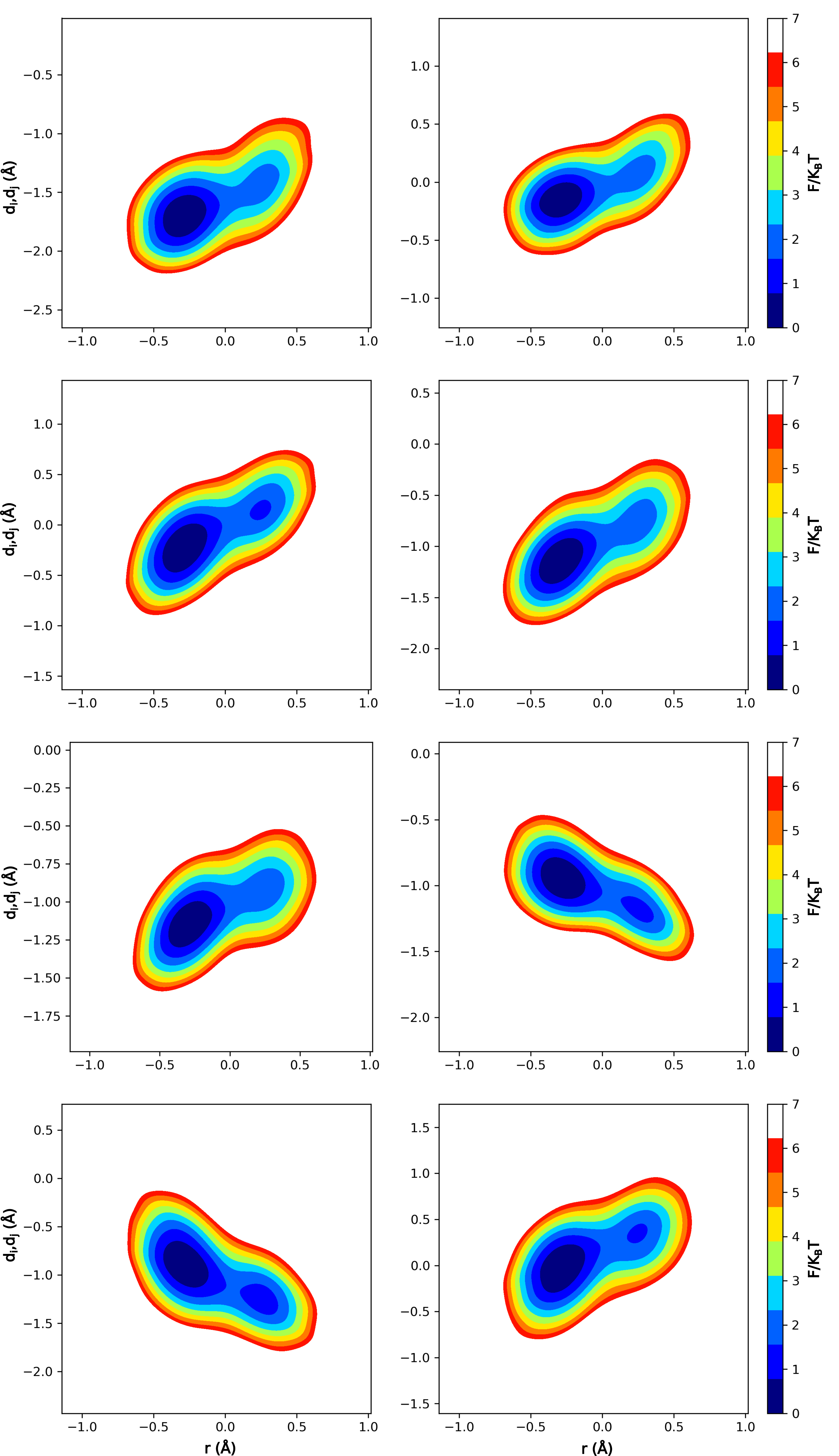}
\caption{
Joint probability distributions between the proton transfer coordinate ($r$) and ammonium rocking modes defined by the distance difference $d_{i,j} = d_i - d_j$, from classical simulations. The plotted modes are selected based on a Pearson correlation coefficient magnitude $|r| > 0.55$ with the PT coordinate.
}
\label{fig:SIcorr}
\end{figure}

Figure~\ref{fig:SIcorr} shows the joint probability distributions between the proton transfer coordinate ($r$) and selected ammonium rocking modes, defined as $d_{i,j} = d_i - d_j$, obtained from classical simulations. The modes shown here correspond to all distance pairs $(i,j)$ for which the absolute value of the Pearson correlation coefficient is larger then 0.55. These distributions highlight the correlated fluctuations between proton position and low-frequency ammonium vibrational modes, consistent with the coupling mechanism discussed in the main text for the representative $d(6,10)$ mode.

\section*{6. Benchmarks for AIMD Simulation Parameters}

\begin{figure}[h!]
    \centering
    \includegraphics[width=0.85\textwidth]{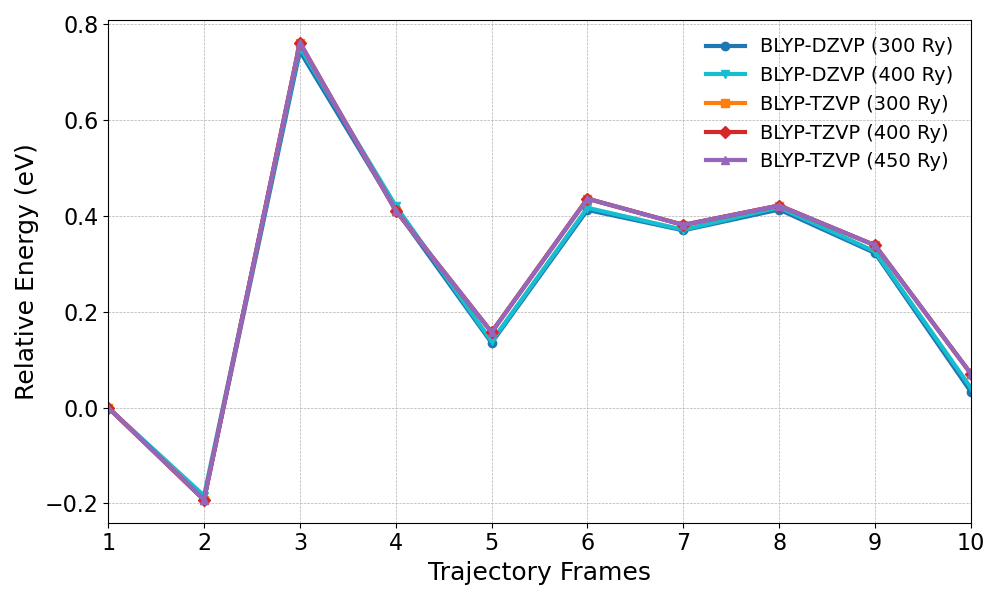}
    \caption{Comparison of single-point energies computed on 10 representative proton-transfer frames using BLYP+D3 with DZVP (300, 400 Ry) and TZVP (300, 400, 450 Ry) settings. All energy differences remain within 0.013 eV.}
    \label{fig:benchmark}
\end{figure}

To validate the choice of basis set and plane-wave cutoff used in the path-integral molecular dynamics (PIMD) simulations, we performed benchmark calculations on 10 representative trajectory frames where a proton transfer event occurs along one of the short hydrogen bonds in the system.
For each frame, single-point total energies were computed using the BLYP functional with D3 dispersion correction, across the following levels of theory:
\begin{itemize}
    \item DZVP basis set with 300 Ry and 400 Ry electronic density cutoffs
    \item TZVP basis set with 300 Ry, 400 Ry, and 450 Ry electronic density cutoffs
\end{itemize}
Figure~\ref{fig:benchmark} shows the relative total energies computed at each level were compared, taking one frame as reference. The maximum deviation across all tested combinations was found to be less than 0.013 eV (13 meV).
These results demonstrate that the DZVP basis set with a 300 Ry cutoff offers a reliable and computationally efficient choice for capturing the relative energetics relevant to proton transfer processes in this system. The use of more demanding TZVP and higher cutoffs did not lead to significant improvements in relative energy accuracy, but would considerably increase the cost of ab initio PIMD simulations.

\end{document}